\documentclass[%
 reprint,
runinaddress,
frontmatterverbose, 
 amsmath,amssymb,
 aps,
 prl,
]{revtex4-2}

\usepackage{graphicx}
\usepackage{dcolumn}
\usepackage{bm}
\usepackage{hyperref}
\usepackage[mathlines]{lineno}
\usepackage{xcolor}
\usepackage[utf8]{inputenc}

\begin{document}

\preprint{APS/123-QED}

\title{Global Framework for Emulation of Nuclear Calculations}

\author{Antoine Belley$^\dagger$}
\email{abelley@mit.edu}
 \affiliation{Massachusetts Institute of Technology, Cambridge, Massachusetts 02139, USA}

\author{Jose M. Munoz$^\dagger$}
\email{josemm@mit.edu}
\affiliation{Massachusetts Institute of Technology, Cambridge, Massachusetts 02139, USA}

\author{Ronald F. Garcia Ruiz}
\email{rgarciar@mit.edu}
\affiliation{Massachusetts Institute of Technology, Cambridge, Massachusetts 02139, USA}
\date{\today}
\collaboration{$^\dagger$\small{Authors contributed equally to this work.}}

\begin{abstract}

We introduce a hierarchical framework that combines ab initio many-body calculations with a Bayesian neural network, developing emulators capable of accurately predicting nuclear properties across isotopic chains simultaneously and being applicable to different regions of the nuclear chart. We benchmark our developments using the oxygen isotopic chain, achieving accurate results for ground-state energies and nuclear charge radii, while providing robust uncertainty quantification. Our framework enables global sensitivity analysis of nuclear binding energies and charge radii with respect to the low-energy constants that describe the nuclear force.
\end{abstract}

\maketitle


\paragraph{\textbf{\textit{Introduction.}---}}The study of nuclear properties and the interactions that govern them is critical to understanding how complex many-body systems such as nuclei, atoms, and molecules emerge from the fundamental forces and particles of nature. At the most fundamental level, the nuclear force is described by Quantum Chromodynamics (QCD), the theory of strong interactions. Due to the non-perturbative complexity of QCD in the low-energy regime, effective interactions derived from Chiral Effective Field Theory ($\chi$EFT)~\cite{Epelbaum:2009,Machleidt:2011} are commonly employed to describe the interactions between neutrons and protons inside the atomic nucleus.
$\chi$EFT provides a low-energy expansion of QCD in terms of nucleon-nucleon and three-nucleon interactions, with undetermined Low-Energy Constants (LECs) adjusted to reproduce few-nucleon observables. However, nuclear properties can be highly sensitive to the value of these LECs~\cite{Hergert:2014, Hebeler:2015, 
Lapoux:2016, Garcia_Ruiz_2016, Hagen:2016, Simonis:2017, Duguet:2017, Lu:2019, Liu:2019, Gysbers:2019, Soma:2020PRC, Belley:2022, Ekström:2023, Arthuis:2024, Belley2024prl}. Therefore, it is crucial to understand the impact of the LECs on nuclear observables and how their uncertainties propagate through \emph{ab initio} many-body calculations.

\emph{Ab initio} methods~\cite{BARRETT:2013,hagen2014coupled, Stroberg:2017, Yao:2020, Soma:2020, Hergert:2020, Tichai:2020, Tichai:2023} rely on systematically improvable approximations to solve the many-body Schrödinger equation, starting from $\chi$EFT interactions.  
Despite major advances in recent years, \emph{ab initio} nuclear calculations of medium and heavy nuclei remain prohibitively expensive due to the exponential scaling of many-body wave functions. While approximate polynomially-scaling methods (e.g.~\cite{Stroberg:2017,hagen2014coupled}) have been developed, this complexity still imposes stringent resource demands and methodological challenges~\cite{Morris:2018,Vernon2022,Karthein2024}. 
Statistical procedures, such as global sensitivity studies, are necessary to evaluate nuclear observables for samples in the range of millions, even for the simplest realistic nuclear interactions ~\cite{Ekstrom:2019}.
As a result, surveying the entire nuclear landscape with realistic uncertainty estimation and establishing a connection to the LECs remains an outstanding challenge.

Nuclear emulators have emerged as powerful tools to address these problems, offering efficient approximations of many-body calculations at substantially reduced computational cost and allowing for uncertainty quantification~\cite{Boehnlein:2022}. Techniques such as multi-output multi-fidelity Deep Gaussian Processes~\cite{Belley2024prl, Belley_thesis}, eigenvector continuation~\cite{Frame:2018,Ekstrom:2019, Drischler2023, Duguet:2024, Jiang:2024} and many-more (e.g.~\cite{Yoshida2023, Giuliani2024, Lay2024}) have been proposed as surrogate models that approximate the outputs of nuclear calculations with significantly reduced computational costs. However, these methods are so far limited to individual isotopes, restricting their ability to capture correlations and trends across isotopic chains and more globally. Moreover, they still demand costly \emph{ab initio} calculation train them.

To overcome the challenges mentioned above, we introduce BAyesian Neural Network for Atomic Nuclei Emulation (BANNANE), a hierarchical Bayesian neural network (BNN) framework that integrates multi-fidelity datasets with categorical and positional embeddings to emulate nuclear properties globally. This approach introduces a flexible architecture tailored for the prediction of different nuclear properties with vastly reduced computational time.
We demonstrate the overarching capabilities of BANNANE by calculating the nuclear binding energies and charge radii of the oxygen isotopic chain, achieving accurate predictions of these observables. Notably, BANNANE enables a global analysis of the sensitivity of these observables to the LECs used in the interactions. This allows us to directly study how macroscopic nuclear properties are impacted by specific details of the nuclear force, and how these sensitivities vary along isotopic chains, which, to our knowledge, is not possible with current nuclear emulators.


\begin{figure*}[t]
    \centering
    \includegraphics[width=0.9\linewidth]{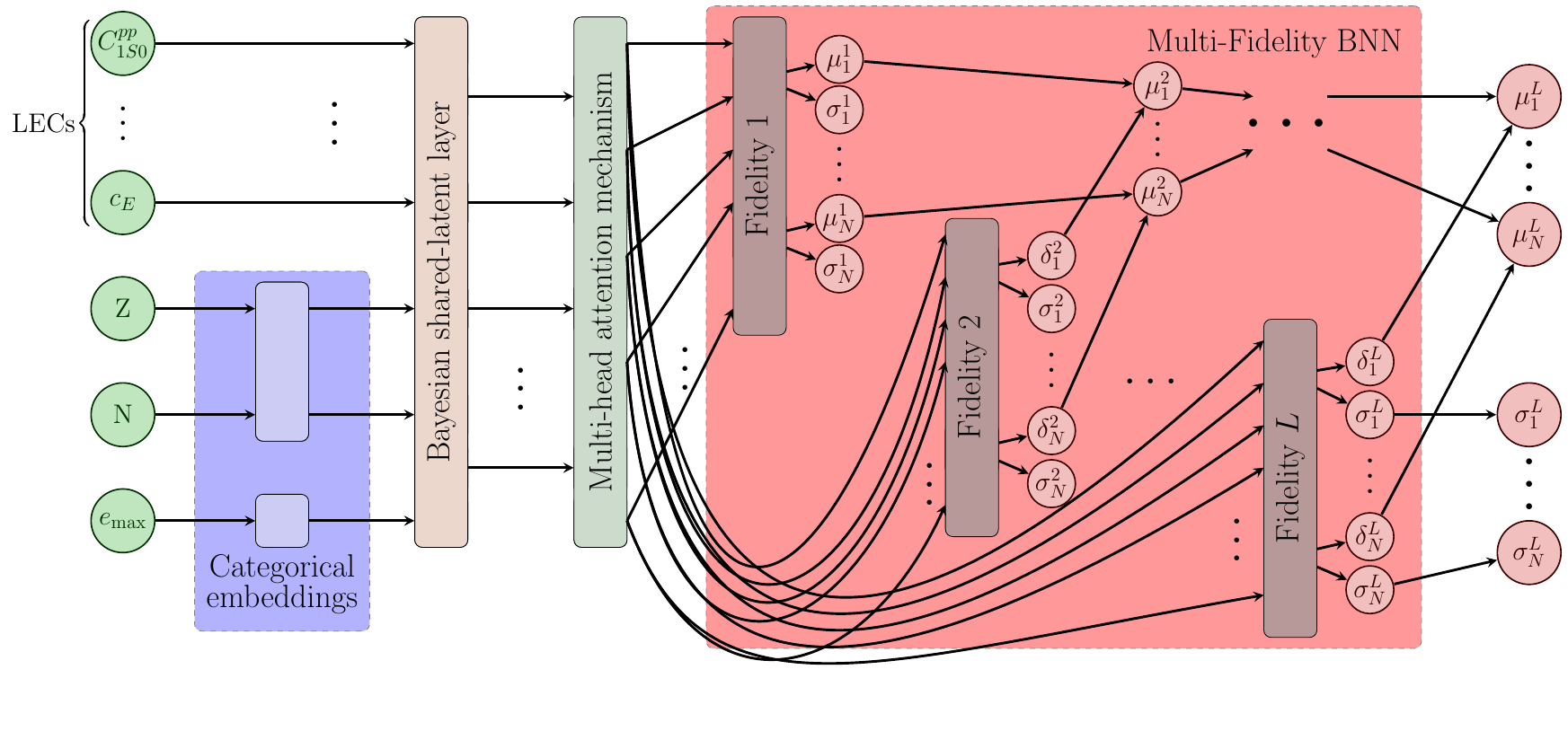}
    \caption{\textbf{BANNANE architecture overview.} The input LECs, along with categorical embeddings for proton number (\(Z\)), neutron number (\(N\)), and fidelity level (\(e_{\text{max}}\)), are processed through a Bayesian shared-latent layer and a multi-head attention mechanism. Fidelity-specific prediction heads refine the base prediction at higher \(e_{\text{max}}\) for each fidelity ranging from 2, outputting mean (\(\mu_N^f\)) and uncertainty (\(\sigma_N^f\)) estimates for nuclear observables.}
    \label{fig:diagram_bannane}
\end{figure*}

\paragraph{\textbf{\textit{Methods.}---}} We construct a multi-fidelity Bayesian emulator using a \emph{hierarchical} architecture tailored for global nuclear modeling. Our pipeline is divided into three main stages: (i) \emph{data loading and preprocessing}, (ii) \emph{hierarchical BNN construction}, and (iii) \emph{variational training and inference}.

\textit{Data generation and preprocessing:} LEC samples utilized in this work are taken from Ref.~\cite{Jiang:2024}, where history matching was used to constrain the LECs to physically plausible ones. More specifically, these samples are for a formulation of $\chi$EFT where $\Delta$-isobars are considered explicitly, up to next-to-next-to leading order (N2LO) in the chiral expansion. In particular, 17 LECs are required to parametrize this theory. We construct data sets for training, validation, and testing randomly from the 8188 samples given in Ref.~\cite{Jiang:2024}.
We employ the Valence-Space formulation of the In-Medium Similarity Renormalization Group (VS-IMSRG)~\cite{Stroberg:2017, Stroberg:2019} to solve the many-body problem and obtain the ground-state energies and charge radii of the different isotopes. In practice, the decoupling of the valence space is done using the \texttt{imsrg++} code~\cite{Stroberg_IMSRG_2018}, and the valence space is then diagonalized using the \texttt{KSHELL} shell-model code~\cite{Shimizu2019}. In particular, we use calculations with different model space sizes, given by including all harmonic oscillator states with $e = 2n+l \leq e_{\rm max}$ where $n$ and $l$ are the principal quantum number and angular momentum of the wave functions respectively. We note that increasing $e_{\rm max}$ results in a more accurate calculation, at the cost of a rapid rise in computational power required to solve the problem. We therefore consider the values of the observables at $e_{\rm max}\in \{4,6,8,10\}$ to constitute different fidelities in the context of the emulator. 

\textit{Hierarchical BNN:} 
Figure~\ref{fig:diagram_bannane} presents a diagrammatic representation of 
the multi-fidelity architecture used in BANNANE. Our core model uses distinct embeddings for the proton number $Z$, and neutron number $N$, via a learnable positional encoding, and $e_{\text{max}}$ of each sample. These embeddings are concatenated with the LECs, then passed through a \emph{shared} multi-head self-attention layer that captures correlations across nuclei and is queried for each fidelity. 
To accommodate multiple fidelities, we combine a \emph{base} predictor (representing the lowest-$e_{\text{max}}$ dataset) with additive \emph{delta} blocks for each higher-$e_{\text{max}}$ level. Concretely, the network first produces a common latent representation, which is mapped to predictions via (i) a base model for the lowest $e_{\text{max}}$ and (ii) incrementally learned offsets for any higher-$e_{\text{max}}$ data present. This ensures consistent cross-$e_{\text{max}}$ learning, allowing the emulator to leverage coarse information from lower-$e_{\text{max}}$ calculations and refine it where higher-$e_{\text{max}}$ data exist.

\textit{Variational training and inference:} We implement the hierarchical BNN with Pyro’s stochastic variational inference (SVI)~\cite{bingham2019pyro}, employing a diagonal Gaussian variational posterior. All weights, biases, and $e_{\text{max}}$-specific parameters (including output variance terms) are learned by minimizing evidence lower bound (ELBO)~\cite{hoffman2016elbo}. After each SVI iteration, the validation loss is monitored for early stopping. During inference, our posterior predictive distribution is sampled for each $(Z, N)$ and $e_{\text{max}}$. This yields both mean predictions and standard deviations, thus quantifying uncertainties across isotopic chains. By design, BANNANE can \emph{extrapolate} to new or sparse regions of the nuclear chart simply by evaluating the learned embeddings of $(Z, N)$. 
Further implementation details and additional ablation studies are provided in the supplemental material~\cite{SupplementalMaterial}.

\begin{figure}[h!]
    \centering
    \includegraphics[width=1
    \linewidth]{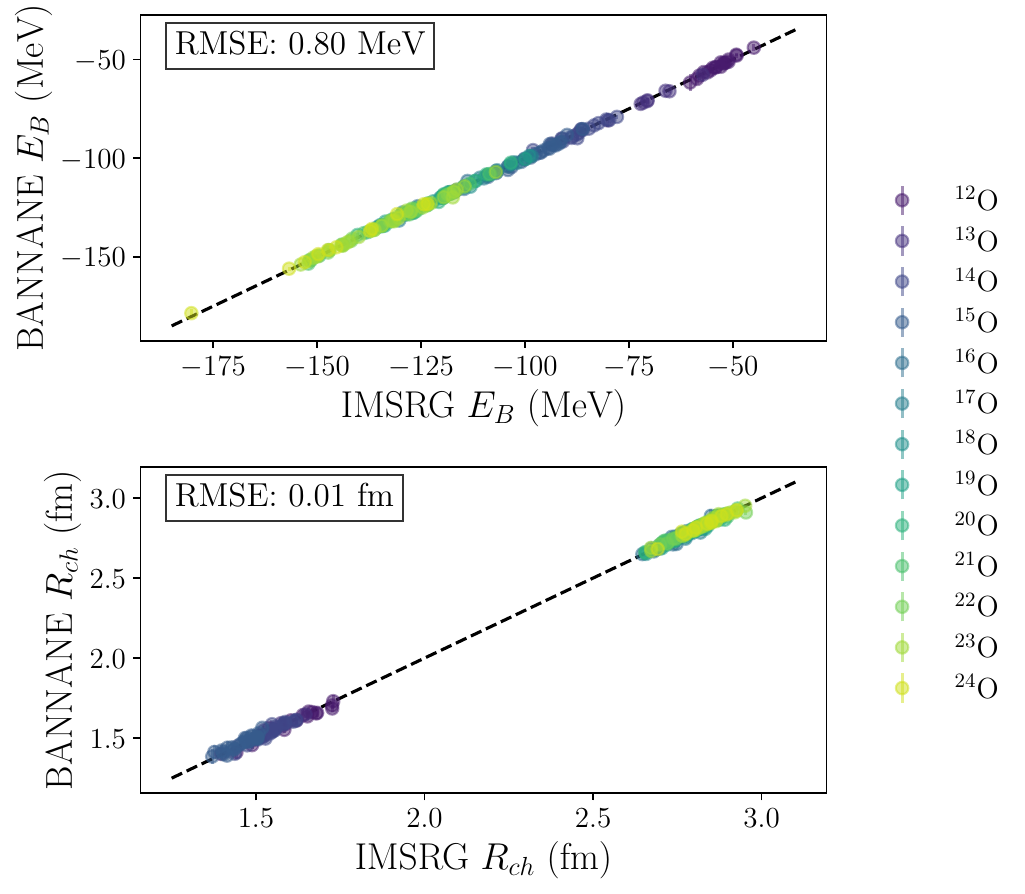}
    \caption{\label{fig:full_oxygen_performance}
    \textbf{Full-chain training on the oxygen isotopes.}
    BANNANE predictions versus IMSRG reference results for test samples at the highest fidelity \(e_{\text{max}}=10\), for \(E_{B}\) and $R_{ch}$. Error bars indicate \(1\sigma\) uncertainty from BANNANE's posterior, and the average Root Mean Squared Error (RMSE) is displayed for each one.
    }
\end{figure}

\paragraph{\textbf{\textit{Results.}---}}In this section, we demonstrate BANNANE’s performance under a range of conditions:

\textit{Performance on the oxygen isotopic chain:} We assess BANNANE on the oxygen isotopic chain ($^{12}$O -- $^{24}$O) including up to $e_{\text{max}}=10$ for all the isotopes. Figure~\ref{fig:full_oxygen_performance} displays the predicted ground-state energies at \( e_{\text{max}}=10\) for all oxygen isotopes, compared to the reference IMSRG calculations. 

We observe that BANNANE correctly reproduces both total binding energies $E_B$ and charge radii $R_{ch}$ across the chain with 0.8 MeV and 0.01 fm Root Mean Squared Error (RMSE), respectively. This represents a significant improvement compared to the Eigenvector Continuation-based subspace-projected coupled-cluster~\cite{Ekstrom:2019} which obtains RMSEs of 3 MeV and 0.02 fm for the ground state energies and charge radii respectively in $^{16}$O using 128 training points. While BANNANE uses more training points in total for this isotope (715  in total), only 51 samples are of the highest fidelity, making BANNANE much more computationally efficient to train, while still achieving an RMSE of 0.337 MeV and 0.02 fm, respectively, for the same isotope.
Further study of the residuals and benchmark is provided in the supplemental material~\cite{SupplementalMaterial}.

\begin{figure}[!h]
    \centering
    \includegraphics[width=1\linewidth]{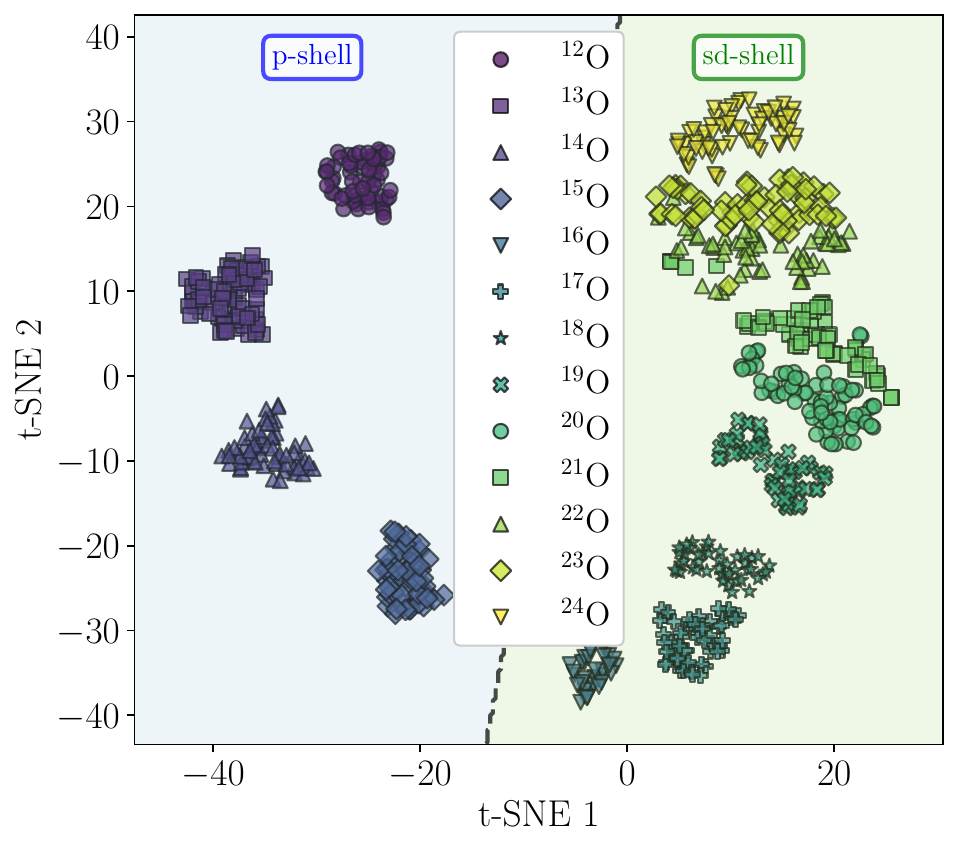}
    \caption{\textbf{Learnt embedding space.} Projection of the attention map to 2 dimensions using t-distributed Stochastic Neighbor Embedding (t-SNE) dimensionality reduction applied to the latent space for test LEC samples. 
    The line and colored background show the decision boundary of a simple linear classifier to the reduced space to illustrate the separation between shells.
    \label{fig:tsne_attention_output_by_N}}
    
\end{figure}

Moreover, we note that the discontinuity caused by the shell closure at $N=8$ is well captured by BANNANE.  To further study this phenomenon, we performed a dimensional reduction to perform an analysis of the latent space of the model after the attention mechanism using a projection t-distributed Stochastic Neighbor Embedding (t-SNE) dimensionality reduction~\cite{van2008visualizing} for visualization. As illustrated in Fig.~\ref{fig:tsne_attention_output_by_N}, there is an evident clustering of the isotopes in the $sd$ shell, after the shell closure at $N=8$, whereas the isotopes in the $p$ shell appear to be mapped to distinct clusters, hinting of the model's ability to capture an underlying nuclear structure.

\textit{Zero-shot extrapolation\label{sec:zero_shot}}: A major advantage of BANNANE is its ability to make reliable predictions for isotopes beyond those included in the training. BANNANE’s architecture enables true zero-shot generalization, meaning it can infer nuclear properties of isotopes it has never seen before, solely from learned trends across isotopic chains.
 To illustrate this, we simultaneously withheld all the data corresponding to $^{15}$O samples and evaluated the model's regression. 

\begin{figure}[!h]
    \centering
    \includegraphics[width=1
    \linewidth]{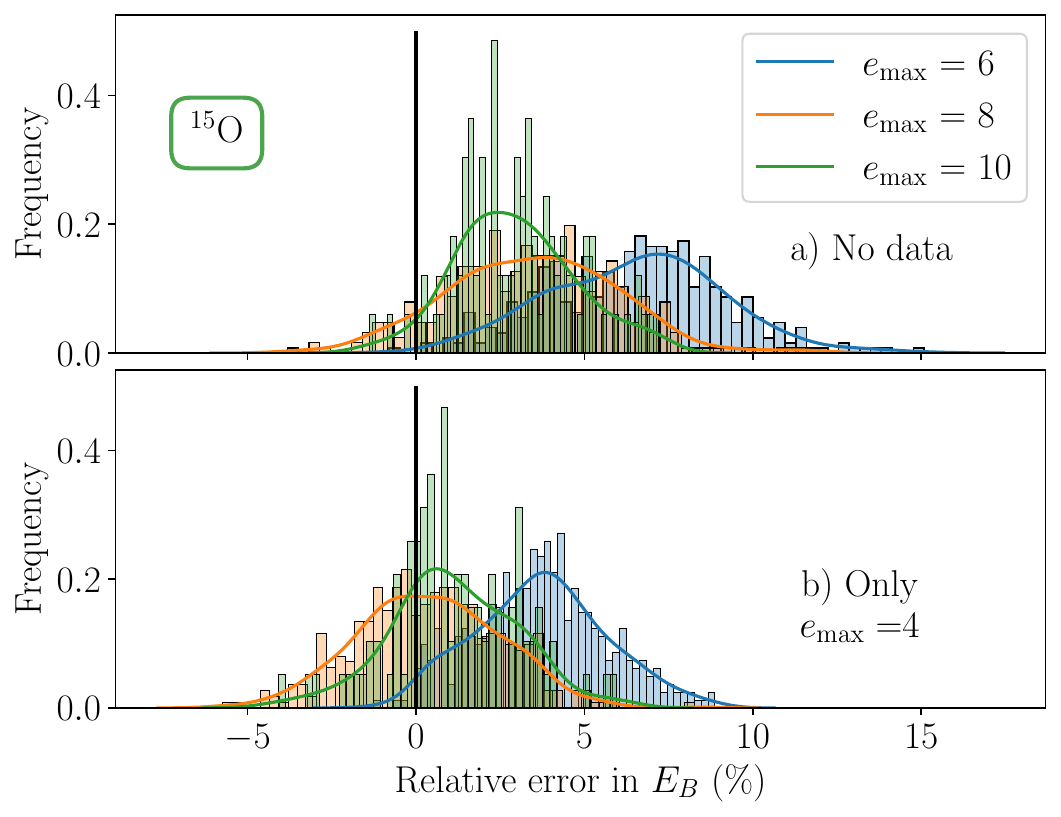} \caption{\textbf{Extrapolation to $^{15}$O.} (a) Residual distribution of the binding energy $E_B$ (\%) for BANNANE predictions compared to IMSRG reference values at $e_{\text{max}} = 6, 8, 10$, when no training samples for $^{15}$O were included. (b) Residual distribution for $E_B$ predictions when only low-fidelity data ($e_{\text{max}} = 4$) was used in training. The inclusion of low-fidelity data significantly reduces systematic bias and improves the overall accuracy of the extrapolated predictions.
\label{fig:zero_shot_interpolation}}
\end{figure}

Figure~\ref{fig:zero_shot_interpolation} (top) shows the relative errors (as a percentage) of $E_B$ once all the samples for this isotope are left out of the training set. Interestingly, the inductive biases on the hierarchical architecture are such that the model can capture both convergence and nuclear structure even for unseen isotopes, as shown by the small residuals at each fidelity. Besides the noticeable shift in the regressed values, it is still surprising that no samples for this isotope were provided. We note that the performance is weaker for the charge radii, especially near the shell closure as discussed along with further analysis for extrapolation over the complete chain in the supplemental material~\cite{SupplementalMaterial}.

This fit can be improved significantly when including data only from the lower fidelity, namely from $e{\rm max} = 4$, as illustrated in Figure~\ref{fig:zero_shot_interpolation} (bottom). We find that including the lower fidelity greatly reduces the spread of the residuals, reduces the systematic bias, and makes the distribution closer to a Gaussian, making for much better-behaved residuals on all fronts.

This extrapolation feature shows great advantages for future applications to heavy systems where high-precision calculations of many LEC samples are infeasible for an entire isotopic chain. In practice, one could choose to compute $e_{\text{max}}=10$ results only for a small \emph{strategic} set of isotopes (e.g., a few in mid-shell regions where higher-order correlations are most relevant and around shell-closure where discontinuity are more likely to happen) and low-fidelity data for the rest of the isotopes. 

Crucially, the Bayesian nature of BANNANE captures the remaining uncertainties due to limited high-fidelity coverage. 
These uncertainty estimates can inform experimental proposals or guide more selective future high-fidelity calculations, focusing effort on regions where the emulator’s confidence is lowest. This phenomenon can be further studied by analyzing how the model's emulation residual converges as more high-fidelity training data is included, for which results are presented in the Result section of the supplemental material~\cite{SupplementalMaterial}.

\begin{figure}[h]
    \centering
    \includegraphics[width=\linewidth]{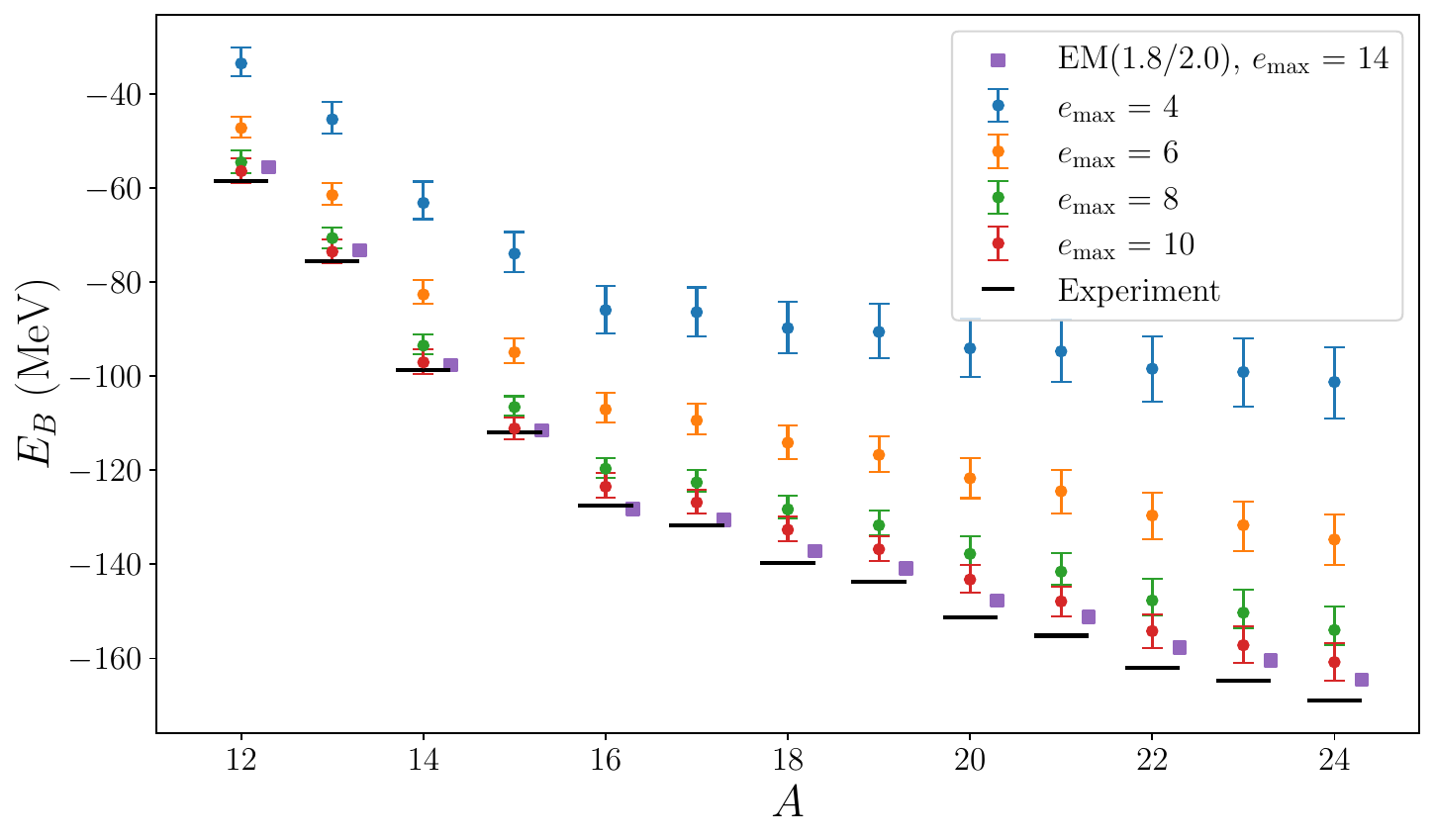}
    \caption{Emulator-driven convergence of binding energies $E_B$ for oxygen isotopes as a function of $e_\text{max}$. Results are compared to fully converged VS-IMSRG results~\cite{Stroberg:2021} at $e_\text{max} =14$ using the EM(1.8/2.0) nuclear interaction in purple squares. Solid lines represent experimental results from \cite{wang2021ame}. \label{fig:energy_predictions}}
\end{figure}
\textit{Physical Convergence:}
BANNANE not only interpolates and extrapolates nuclear observables but also emulates the physical convergence of many-body methods. Figure~\ref{fig:energy_predictions} shows the emulator's predictions for \(E_B\) across the oxygen chain as a function of \(e_{\text{max}}\), compared to experimental values after doing a weighted resampling of 8188 LEC samples~\cite{Jiang:2024} as done in Ref.~\cite{Belley2024prl}. The emulator accurately reproduces convergence trends, including nontrivial shell effects, with deviations well within its uncertainty estimates, and in good agreement with previous results~\cite{Lapoux_2016}. Notably, these uncertainties stem solely from the emulator and the LECs, not from many-body truncation effects. This highlights BANNANE’s ability to capture systematic trends while reducing computational costs, which in practice could take several years on HPC clusters to just a few seconds for the resampling. 

\begin{figure*}[t!]
    \centering
    \includegraphics[width=1\linewidth]{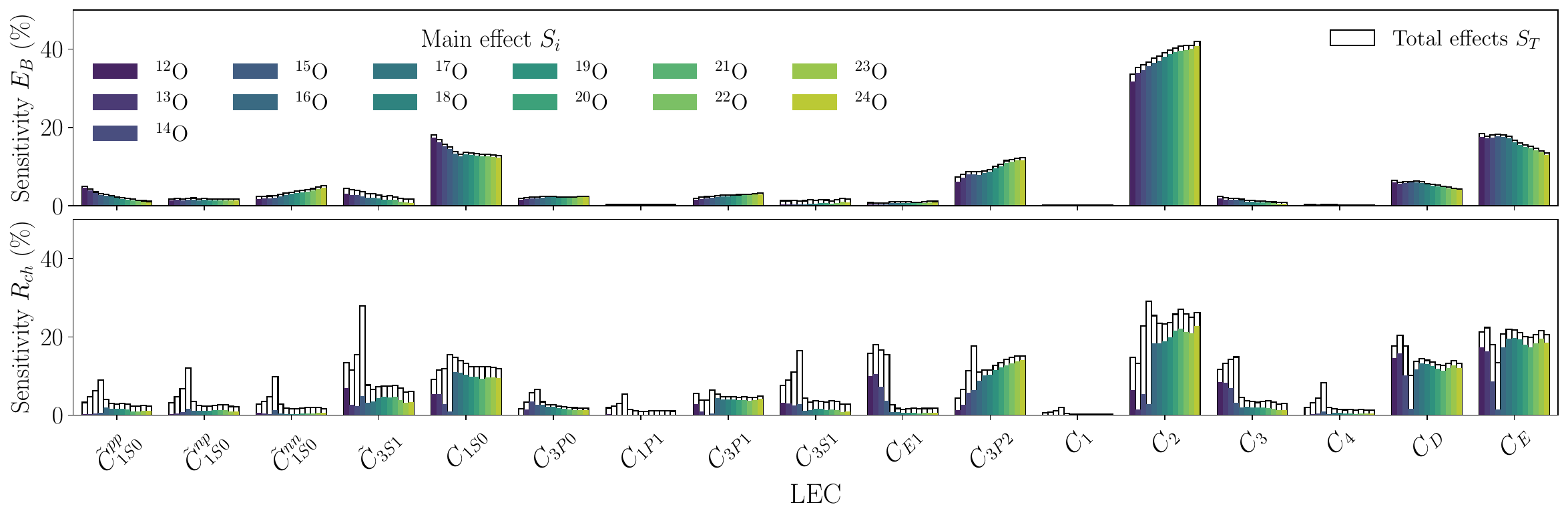}
    \caption{\textbf{Variance-based global sensitivity analysis for the oxygen isotopes.} 
Shown are the main-effect indices $S_i$ (colored bars) and total-effect indices $S_T$ (black outline) for each LEC, indicating their direct and combined contributions to the variance of the binding energy $E_B$ (top) and charge radius $R_{ch}$ (bottom). Notable differences between $S_i$ and $S_T$ point to strong interactions among particular LECs.
}
    \label{fig:sobols}
\end{figure*}

\textit{Unified Sensitivity Estimation:}
Beyond providing fast and accurate emulation, BANNANE enables global sensitivity analysis (GSA) that would be computationally prohibitive using brute-force \emph{ab initio} methods. With this method, we can probe how much the variance of each LEC (and their covariance) impacts the variance of the observables.
In Fig.~\ref{fig:sobols}, we apply a variance-based Sobol approach~\cite{sobol2001global} to assess how each LEC affects $E_B$ and $R_{ch}$ across oxygen isotopes from $^{12}$O to $^{24}$O. 
We plot the main-effect index $S_i$ (filled bars), capturing the direct contribution of each LEC to the variance of an observable, and the total-effect index $S_T$ (white outlines), accounting for higher-order interactions (see supplemental material~\cite{SupplementalMaterial} for more details.).

A few features stand out: For binding energies (top panel), certain two-nucleon (2N) couplings (e.g., $C_{1S0}$ and $c_2$) and three-nucleon (3N) couplings (e.g., $C_D$, $C_E$) dominate the variance while showing small progressive changes across the chain. This is consistent with findings of previous methods~\cite{Ekstrom:2019, Ekström:2023,Belley2024prl} in multiple nuclei. It indicates that the dependency of the ground state energy to the LEC exhibits small variations in different nuclear systems. In contrast, the hierarchy of LEC sensitivities differs considerably for the charge radii (bottom panel): large changes can be observed at the shell-closure, with for example $C_{3P2}$ being important and $C_{E1}$ being negligible in the sd-shell but the opposite holding true in the p-shell. Moreover, the sensitivities for $R_{ch}$ are inherently more nonlinear and show the emerging importance of cross-couplings rather than individual influences. This highlights that the charge radius can be highly sensitive to certain LECs, making it strongly complementary to the experimental constraints derived from binding energies. In particular, experimental data on neutron-deficient oxygen isotopes, which have not yet been measured, will be crucial for guiding these developments and benchmarking our predictions.

\paragraph{\textbf{\textit{Conclusions.}---}}

We have introduced a hierarchical multi-fidelity Bayesian Neural Network for Atomic Nuclei Emulation (\textbf{BANNANE}), which synthesizes low- and high-fidelity datasets across different nuclei. By leveraging learnable embeddings for nucleon numbers and an additive fidelity-specific architecture, BANNANE not only interpolates faithfully among known nuclei, achieving better accuracy than current emulators but also extrapolates with remarkable accuracy to hitherto uncalculated (or unobserved) nuclei. Due to this, BANNANE offers a computationally feasible way to calculate nuclear properties with realistic uncertainties and with an accuracy comparable to those obtained by considerably costly \emph{ab initio} methods. These developments have enabled us to investigate the overall dependence of nuclear observables on the LECs that govern inter-nucleon interactions. We found that the nuclear binding energy and charge radii of oxygen isotopes exhibit distinct sensitivities to these LECs, with particular interest in data from neutron-deficient oxygen isotopes that have yet to be measured. We hope our findings will motivate charge radius measurements of these isotopes, which are now feasible via laser spectroscopy experiments \cite{Yan23} at the new Facility for Rare Isotope Beams (FRIB) \cite{FRIB25}. More broadly, we anticipate that BANNANE will serve as a powerful tool to guide future experiments and establish direct connections between measurements of binding energy, charge radii, and specific components of the nuclear force.

Beyond its computational advantages, BANNANE provides a powerful framework to guide future experimental and theoretical efforts. It enables targeted high-precision calculations and experimental searches where they matter most, optimizing resource allocation for both theory and experiments. Its versatile architecture can be used with any many-body method and can include as many observables as desired. While we have only varied the neutron number in this study, BANNANE architecture also allows the emulation of different proton numbers, which we will test in future work.

\paragraph{\textbf{\textit{Data Availability.}---}}
The data for reproducing the results in this work can be found online~\cite{munozariasjm_paper_o_bannane} along with scripts used in the data analysis. Additionally, the source code for training the model is available upon request.

\begin{acknowledgments}

We thank A. Ekström, C. Forssén, G. Hagen, and W. G. Jiang for providing the interaction samples used in this work, and J. Holt for insightful discussions. The IMSRG code used in this work makes use of the Armadillo \texttt{C++} library \cite{Sanderson2016, Sanderson2018}.  Computational resources were provided by subMIT at MIT Physics. This work was supported by the Office of Nuclear Physics, U.S. Department of Energy, under grants DESC0021176 and DE-SC0021179. We acknowledge the support of the Natural Sciences and Engineering Research Council of Canada (NSERC) [PDF-587464-2024].

\end{acknowledgments}

\bibliography{main_arxiv}

\clearpage
\newpage
\section{Supplemental Material}

\section{Methods}
\subsection{Architecture Details\label{sec:SI_arch}}

BANNANE is designed to emulate nuclear properties across isotopic chains while integrating multi-fidelity data. Its hierarchical Bayesian framework combines learnable embeddings, a multi-head self-attention mechanism with fidelity-specific queries, and fidelity-specific prediction pathways. This architecture ensures that the emulator can leverage correlations between nuclear observables across elements and isotopes, while also adapting to varying levels of computational fidelity. It must be added that we expect the model to further improve the performance once a full hyperparameter optimization is performed. Below we present a detailed overview of our implementation.

\paragraph{\textbf{\textit{Input embeddings.}---}}
To incorporate discrete nuclear identifiers and fidelity information, BANNANE utilizes three primary embedding layers:

\paragraph{Element (\(Z\)) embedding:}
An embedding layer maps each proton number \(Z \in \{0, 1, \ldots, Z_{\text{max}}\}\) to a continuous vector space:
\[
\mathbf{e}_Z = \mathbf{W}_Z \cdot \mathbf{1}_Z,
\]
where \(\mathbf{W}_Z \in \mathbb{R}^{(Z_{\text{max}}+1) \times d_Z}\) is a learnable weight matrix, \(\mathbf{1}_Z\) is a one-hot encoding of \(Z\), and \(d_Z\) is the embedding dimension. To increase the architecture's flexibility, we left the option to include this embedding into the model, which is not considered for the benchmarks presented in this work but will be needed when other $Z$ values are considered.

\paragraph{Neutron number (\(N\)) positional encoding:}
A similar embedding can be applied to the neutron number \(N\): $\mathbf{e}_N = \mathbf{W}_N \cdot \mathbf{1}_N$ with \(\mathbf{W}_N \in \mathbb{R}^{(N_{\text{max}}+1) \times d_N}\) and \(d_N\) the corresponding dimension. This learned positional encoding allows the network to capture trends as a function of neutron number, crucial for modeling isotopic evolution.

However, instead of a learnable embedding, BANNANE uses a fixed sinusoidal positional encoding inspired by Transformer architectures. This encoding defines a phase:
$$
\varphi = \cdot \exp\left(-\frac{2i}{d_N}\log(10000)\right), \quad i = 0, \dots, \left\lfloor\frac{d_N-2}{2}\right\rfloor
$$

Which is then added as a positional encoding via:

\[
\begin{aligned}
\text{PE}(n, 2i) = \sin\left(n\varphi)\right),\quad
\text{PE}(n, 2i+1) = \cos\left(n \varphi\right).
\end{aligned}
\]

This guarantees that the neutron number encoding captures smooth trends across isotopic chains, aiding generalization. The sinusoidal encoding is implemented as a configurable alternative to learned embeddings, selectable via hyperparameters. We used the sinusoidal encoded $N$ embeddings for all the results in this work.

\paragraph{Fidelity level embedding:}
Each fidelity level \(f \in \mathcal{F}\) is similarly embedded:
\[
\mathbf{e}_f = \mathbf{W}_f \cdot \mathbf{1}_f,
\]
where \(\mathbf{W}_f \in \mathbb{R}^{|\mathcal{F}| \times d_f}\) maps a one-hot representation of fidelity level to an embedding space of dimension \(d_f\).

The continuous input features \(\mathbf{x} \in \mathbb{R}^{d_x}\), which may include coupling constants and other parameters derived from effective field theories, are concatenated with these embeddings to form a combined input:
\[
\mathbf{z} = \Big[ \mathbf{x},\, \mathbf{e}_Z,\, \mathbf{e}_N,\, \mathbf{e}_f \Big] \in \mathbb{R}^{d_x + d_Z + d_N + d_f}.
\]

\paragraph{\textbf{\textit{Shared latent representation.}---}}
The combined input \(\mathbf{z}\) is fed into a \emph{shared latent network} to extract a feature-rich representation:
\[
\mathbf{h} = \phi(W_2 \cdot \phi(W_1 \cdot \mathbf{z} + \mathbf{b}_1) + \mathbf{b}_2),
\]
where \(W_1\), \(W_2\) are weight matrices, \(\mathbf{b}_1\), \(\mathbf{b}_2\) are biases, and \(\phi(\cdot)\) denotes a non-linear activation function (e.g., LeakyReLU). This produces a latent vector \(\mathbf{h} \in \mathbb{R}^{d_h}\) shared across all fidelities.

\paragraph{\textbf{\textit{Multi-head self-attention mechanism with fidelity-specific queries.}---}}
To capture complex interdependencies and fidelity-specific features, BANNANE employs a multi-head self-attention mechanism augmented with fidelity-specific query vectors~\cite{vaswani2017attention}. This allows the model to extract different aspects of the shared representation tailored to each fidelity level. For our implementation, we fixed the number of attention heads to 2.

\paragraph{Fidelity-specific queries:}  
For each fidelity level, including the base fidelity, a unique learnable query vector is introduced:
\[
\mathbf{Q}^{(f)} = \mathbf{W}_q^{(f)} \cdot \mathbf{h},
\]
where \(\mathbf{W}_q^{(f)}\) is a learnable projection matrix specific to fidelity \(f\).

\paragraph{Attention computation:}  
For a given fidelity \(f\), the query \(\mathbf{Q}^{(f)}\) interacts with the shared latent representation \(\mathbf{h}\) to compute attention weights and extract relevant features:
\[
\text{Attention}^{(f)}(\mathbf{Q}^{(f)}, \mathbf{K}, \mathbf{V}) = \mathrm{softmax}\left(\frac{\mathbf{Q}^{(f)} \mathbf{K}^\top}{\sqrt{d_k}}\right) \mathbf{V},
\]
where \(\mathbf{K} = W_k \mathbf{h}\) and \(\mathbf{V} = W_v \mathbf{h}\) are shared key and value projections, respectively, and \(d_k\) is the dimension per head.

\paragraph{Concatenation and projection:}  
The outputs from all attention heads are concatenated and passed through a linear layer to form the final attention-enhanced representation:
\[
\mathbf{o}^{(f)} = W_o \Big[ \text{head}_1^{(f)}, \ldots, \text{head}_H^{(f)} \Big],
\]
with \(W_o \in \mathbb{R}^{(H \cdot d_k) \times d_{\text{out}}}\).

\paragraph{\textbf{\textit{Fidelity-specific regression pathways.}---}}
The attention-enhanced representation \(\mathbf{o}^{(f)}\) is used to predict target observables hierarchically, leveraging both shared and fidelity-specific information:

\paragraph{Base prediction:}
For the lowest fidelity level \(f_0\), the model computes
\[
\begin{aligned}
    \mathbf{y}^{(f_0)} &= f_{\text{base}}(\mathbf{o}^{(f_0)})\\ 
    &= W_{\text{base}}^2 \cdot \phi(W_{\text{base}}^1 \cdot \mathbf{o}^{(f_0)} + \mathbf{b}_{\text{base},1}) + \mathbf{b}_{\text{base},2},
\end{aligned}
\]
where \(W_{\text{base}}^1\), \(W_{\text{base}}^2\) and biases \(\mathbf{b}_{\text{base},1}, \mathbf{b}_{\text{base},2}\) are learnable parameters.

\paragraph{Delta adjustments for higher fidelities:}
For each subsequent fidelity level \(f > f_0\), the model refines the prediction by adding a delta adjustment:
\[
\Delta \mathbf{y}^{(f)} = f_{\text{delta}, f}(\mathbf{o}^{(f)}) = W_{f}^2 \cdot \phi(W_{f}^1 \cdot \mathbf{o}^{(f)} + \mathbf{b}_{f,1}) + \mathbf{b}_{f,2}.
\]
The prediction at fidelity \(f\) is then given by the sum of the base prediction and all applicable delta corrections:
\[
\mathbf{y}^{(f)} = \mathbf{y}^{(f_0)} + \sum_{f' \leq f} \Delta \mathbf{y}^{(f')}.
\]
In this hierarchical structure, the shared attention mechanism with fidelity-specific queries enables each delta model to focus on different features of the shared representation, enhancing the model's ability to capture fidelity-dependent nuances.

\subsection{Bayesian Inference and Variational Training}

\textbf{\textit{Probabilistic model specification.}---}
The Bayesian neural network (BNN) places a prior distribution over all its weights \(\theta\) and output noise standard deviations \(\sigma\). For instance, we may use a half-normal prior for \(\sigma\):
\[
p(\sigma) = \mathrm{HalfNormal}(\sigma \mid 1.0),
\]
and standard normal priors for weights:
\[
p(\theta) = \prod_{i} \mathcal{N}(\theta_i \mid 0, 1).
\]

Given input \(\mathbf{x}\), fidelity index \(f\), \(Z\), and \(N\), the likelihood of observing target \(\mathbf{y}\) is
\[
p(\mathbf{y} \mid \mathbf{x}, f, Z, N, \theta, \sigma) = \mathcal{N}\Big(\mathbf{y}; \, \mu(\mathbf{x}, f, Z, N; \theta),\, \operatorname{diag}(\sigma^2)\Big),
\]
where \(\mu(\cdot)\) is the output of the hierarchical network.

\textbf{\textit{Variational qpproximation and ELBO.}---}
We approximate the intractable posterior \(p(\theta, \sigma \mid \mathcal{D})\) with a variational distribution \(q(\theta, \sigma)\), typically chosen as a mean-field Gaussian:
\[
q(\theta, \sigma) = \prod_{i} \mathcal{N}(\theta_i \mid m_i, s_i^2) \times q(\sigma),
\]
where \(m_i\) and \(s_i\) are variational parameters. For \(\sigma\), we similarly choose a tractable form.

The Evidence Lower Bound (ELBO) over the dataset \(\mathcal{D}\) is
\[
\begin{aligned}
\mathcal{L}(q) &= \mathbb{E}_{q(\theta, \sigma)}\Big[ \sum_{(\mathbf{x}, \mathbf{y}, f, Z, N) \in \mathcal{D}} \log p(\mathbf{y} \mid \mathbf{x}, f, Z, N, \theta, \sigma) \Big] \\
&\quad - \mathrm{KL}\Big(q(\theta, \sigma) \,\Big\|\, p(\theta, \sigma)\Big).
\end{aligned}
\]

\paragraph{\textbf{\textit{Loss components.}---}}
For a mini-batch \(\mathcal{B}\) sampled from fidelity level \(f\), the negative ELBO loss function decomposes as:
\[
\mathcal{L}_{\mathcal{B}} = \underbrace{-\sum_{i \in \mathcal{B}} \log p(\mathbf{y}_i \mid \mathbf{x}_i, f, Z_i, N_i, \theta, \sigma)}_{\text{Reconstruction Loss}} + \underbrace{\mathrm{KL}\Big(q(\theta, \sigma) \,\Big\|\, p(\theta, \sigma)\Big)}_{\text{Regularization}}.
\]

Optionally, a contrastive loss term \(\mathcal{L}_{\text{contrastive}}\) can be added to encourage similar embeddings for similar nuclei or fidelities:
\[
\mathcal{L}_{\text{total}} = \mathcal{L}_{\mathcal{B}} + \lambda \,\mathcal{L}_{\text{contrastive}},
\]
with \(\lambda\) a hyperparameter weighting the contrastive component.

\subsection{Training Protocol\label{sec:SI_training}}

The training procedure involves iteratively updating the variational parameters to maximize the ELBO:

\begin{enumerate}
    \item \textbf{Initialization:}
    \begin{itemize}
        \item Set random seeds for reproducibility.
        \item Initialize network weights, biases, and variational parameters.
        \item Load multi-fidelity isotopic data, then preprocess and split into training, validation, and test sets.
        \item Standardize input features \(\mathbf{x}\) and target variables \(\mathbf{y}\) using global scalers.
    \end{itemize}
    \item \textbf{Stochastic Variational Inference Loop:}
    \begin{itemize}
        \item Sample mini-batches \(\mathcal{B}_f\) from each fidelity level \(f\).
        \item Compute gradients of the negative ELBO \(\mathcal{L}_{\mathcal{B}_f}\) with respect to variational parameters.
        \item Update parameters using an optimizer like ClippedAdam.
    \end{itemize}
    \item \textbf{Early Stopping and Checkpointing:}
    \begin{itemize}
        \item Track the validation loss and apply early stopping based on patience.
        \item Save the model corresponding to the lowest validation loss.
    \end{itemize}
\end{enumerate}

\begin{figure}
    \centering
    \includegraphics[width=1\linewidth]{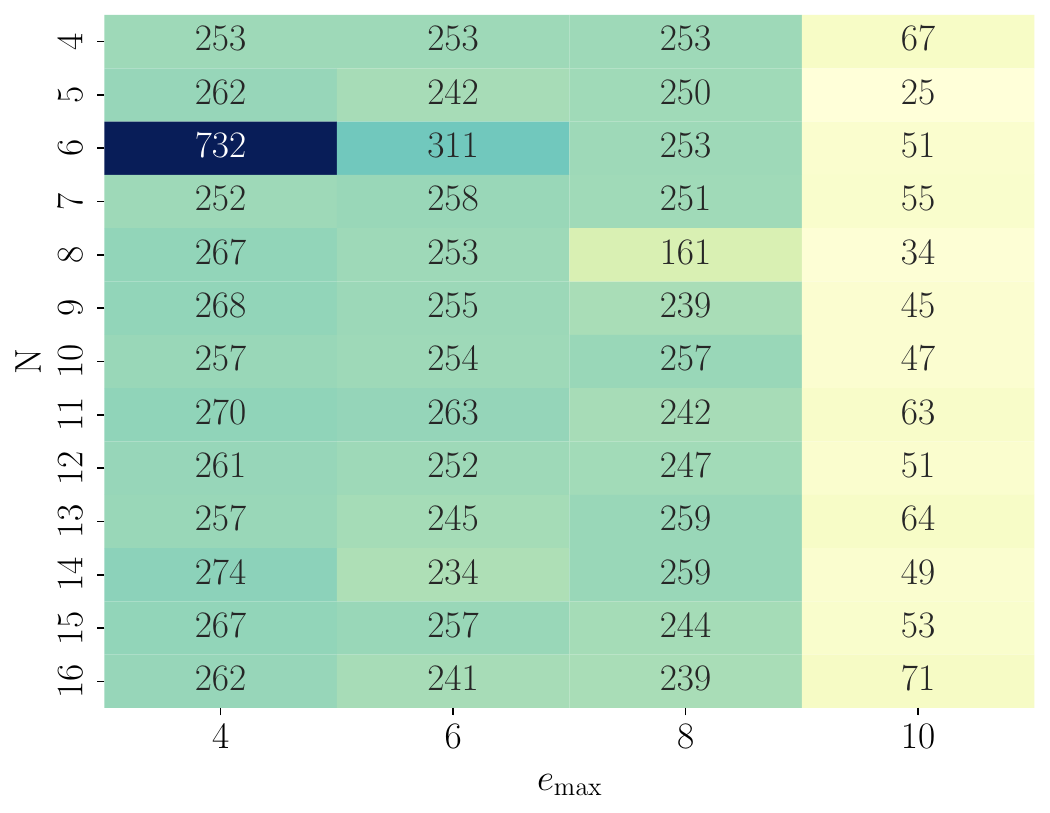}
    \caption{\textbf{Samples per fidelity:} Number of LEC samples per fidelity used for training BANNANE for each one of the isotopes.}
    \label{fig:train_num_heatmap}
\end{figure}

The number of samples per each one of the fidelities and isotopes used in the training procedure is shown in Fig.~\ref{fig:train_num_heatmap}.

\subsection{Inference and Uncertainty Quantification}

To predict for a new input \((\mathbf{x}^*, f^*, Z^*, N^*)\), we sample from the posterior predictive distribution following:

\begin{enumerate}
    \item \textbf{Preprocessing:} Scale \(\mathbf{x}^*\) using the fitted StandardScaler.
    \item \textbf{Monte Carlo Sampling:} Draw \(S\) samples \(\{(\theta^{(s)}, \sigma^{(s)})\}_{s=1}^S\) and compute predictions:
    \[
    \mathbf{y}^{*(s)} = \mu(\mathbf{x}^*, f^*, Z^*, N^*; \theta^{(s)}).
    \]
    \item \textbf{Estimating Statistics:}
    \[
    \hat{\mathbf{y}}^* = \frac{1}{S} \sum_{s=1}^S \mathbf{y}^{*(s)}, \quad \hat{\sigma}^* = \sqrt{\frac{1}{S-1} \sum_{s=1}^S \Big( \mathbf{y}^{*(s)} - \hat{\mathbf{y}}^* \Big)^2}.
    \]
\end{enumerate}

\subsection{Hyperparameters}
To achieve the results presented in this work, we used a shared latent dimension of 80 with hidden layers of 64 units and a multi-head attention mechanism with 2 heads. The neutron number was encoded using a sinusoidal encoding, while the fidelity levels were embedded in an 8-dimensional space. The model included a dropout rate of 0.05 to enhance generalization. Training was performed using the ClippedAdam optimizer with an initial learning rate of \(1\times10^{-4}\), and a patience of 200 for early stopping. The learning rate was reduced by a factor of 2 if validation loss did not improve within 25 evaluations. The number of training iterations was set to 30,000, with validation and test splits of 20\% each. To account for the varying importance of different fidelities, we applied weighted losses, with higher fidelity levels receiving greater weight (\(e_{\text{max}}=4\) to \(10\) assigned weights from 1.0 to 2.5). 

With this configuration for the architecture, the total number of trainable parameters of the architecture is $2\cdot 10^5$.

\section{Results}

\subsection{Complete Oxygen Chain}
To assess the benefits of the global fitting performed by BANNANE, we compare the global fitting with a local fit performed using a state-of-the-art Boosted Decision Trees (BDT)~\cite{prokhorenkova2018catboost} trained on a per-isotope basis including all fidelities, where the fidelity was included as an extra categorical feature. Table~\ref{tab:per_isotope_rmse_comparison} presents a comparison of the global Root-mean-square error (RMSE) per isotope for both approaches.
\begin{table}[h!]
    \caption{\label{tab:per_isotope_rmse_comparison} 
    Root-mean-square error (RMSE) of binding energy ($E_B$) and charge radius ($R_{ch}$) for oxygen isotopes at $e_{\text{max}}=10$.}
    \centering
    \setlength{\tabcolsep}{6pt} 
    \renewcommand{\arraystretch}{1.1} 
    \begin{tabular}{c|cc|cc}
        \hline\hline
        \textbf{N} & \multicolumn{2}{c|}{\textbf{BANNANE}} & \multicolumn{2}{c}{\textbf{BDT}} \\
        \cline{2-5}
        & $E_B$ (MeV) & $R_{ch}$ (fm) & $E_B$ (MeV) & $R_{ch}$ (fm) \\
        \hline
         4  & 0.692(110) & 0.0318(8)  & 2.521(89) & 0.0494(3) \\
         5  & 0.415(72)  & 0.0248(11) & 3.256(14) & 0.0498(2) \\
         6  & 0.642(14) & 0.0131(3)  & 4.541(90) & 0.0285(1) \\
         7  & 0.544(12) & 0.0136(2)  & 3.880(16) & 0.0240(1) \\
         8  & 0.337(55)  & 0.0205(7)  & 5.535(15) & 0.0300(1) \\
         9  & 0.317(45)  & 0.0082(2)  & 3.619(18) & 0.0239(1) \\
        10  & 0.551(11) & 0.0116(2)  & 5.294(11) & 0.0339(1) \\
        12  & 0.455(66)  & 0.0106(2)  & 4.892(31) & 0.0246(1) \\
        13  & 0.293(32)  & 0.0083(1)  & 8.109(30) & 0.0409(1) \\
        14  & 0.277(32)  & 0.0199(5)  & 7.931(44) & 0.0450(2) \\
        15  & 0.385(57)  & 0.0101(1)  & 6.318(24) & 0.0364(1) \\
        16  & 0.538(11) & 0.0087(1)  & 5.182(27) & 0.0340(1) \\
        \hline\hline
    \end{tabular}
\end{table}

\subsection{Leave One Out - Zero Shot extrapolation\label{sec:si_zero_shot}}
\begin{figure}[t]
    \centering
    \includegraphics[width=1.1\linewidth]{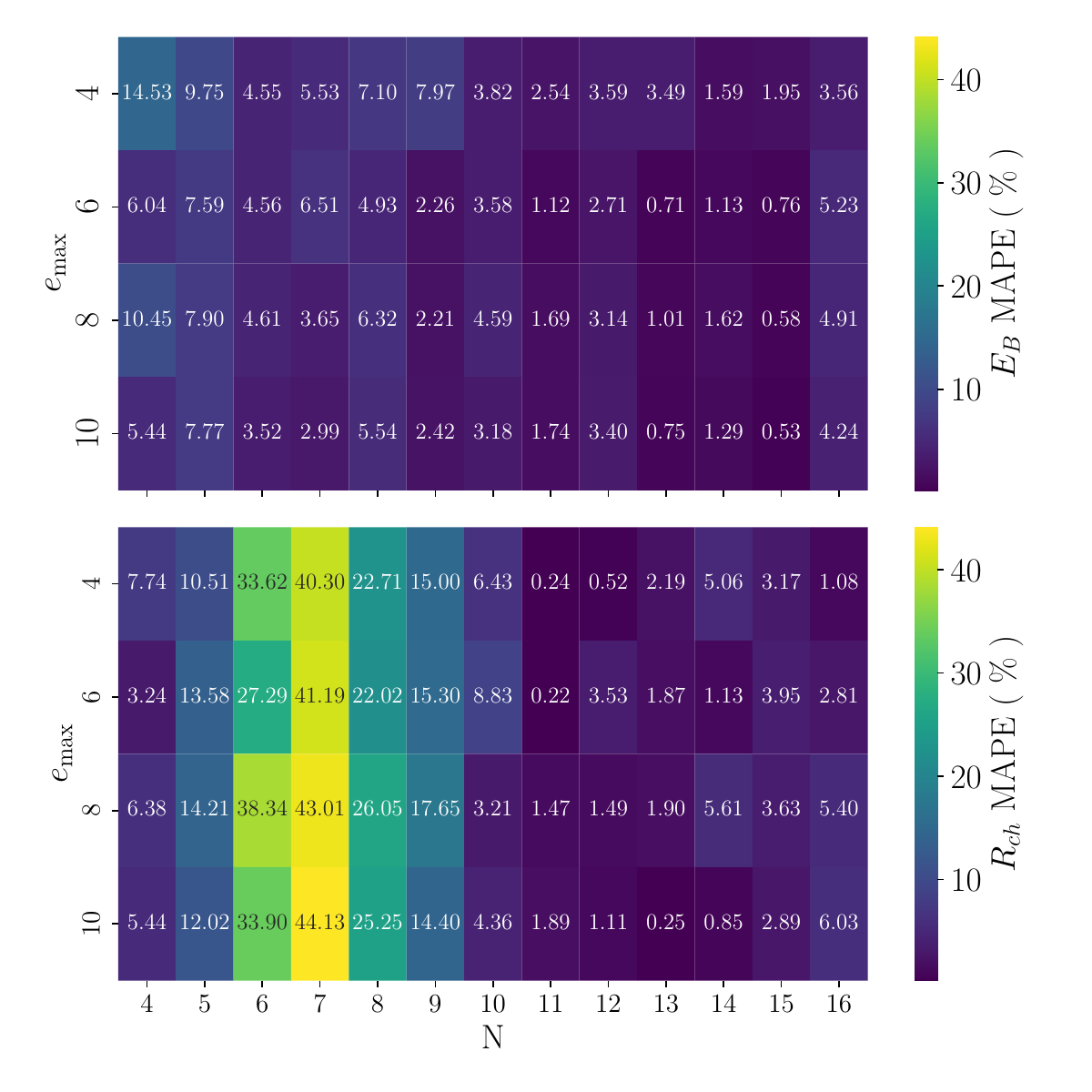}
    \caption{Zero-Shot Leave-On-Out. Distribution of the Zero-Shot extrapolation for unseen isotopes as each isotope was removed from the training Samples.\label{fig:loo_mape_heatmap}}
\end{figure}

As a tool to probe the extrapolation capabilities of BANNANE, we proceeded to remove completely each one of the isotopes from the training samples, and then we proceeded to use the model to evaluate the Mean Absolute Percentage Error (MAPE) for each one of them in a Zero-Shot approach, meaning that no examples for this fidelity were provided before. The distribution of residuals is shown in Fig~\ref{fig:loo_mape_heatmap}. Remarkably, the model does not diverge significantly near the driplines, and the MAPEs do not increase considerably for higher $e_{\text{max}}$ due to self-consistency from the hierarchical architecture.

We find however that the model struggles with identifying the position of the nuclear shell closure if it isn't included in the training, as seen by the significant error in the charge radii when removing one isotope near the edge of the shells. This means that it is important to add nuclei near the edges of the shell for the model to perform well. This is luckily not an issue for many-body methods as these nuclei are simpler to compute. Nuclear shells could be given as an input of the model in the future to remedy this.

\subsection{Fidelity Extrapolation\label{sec:si_fidelity_extra}}

\begin{figure}[t]
    \centering
    \includegraphics[width=1.1\linewidth]{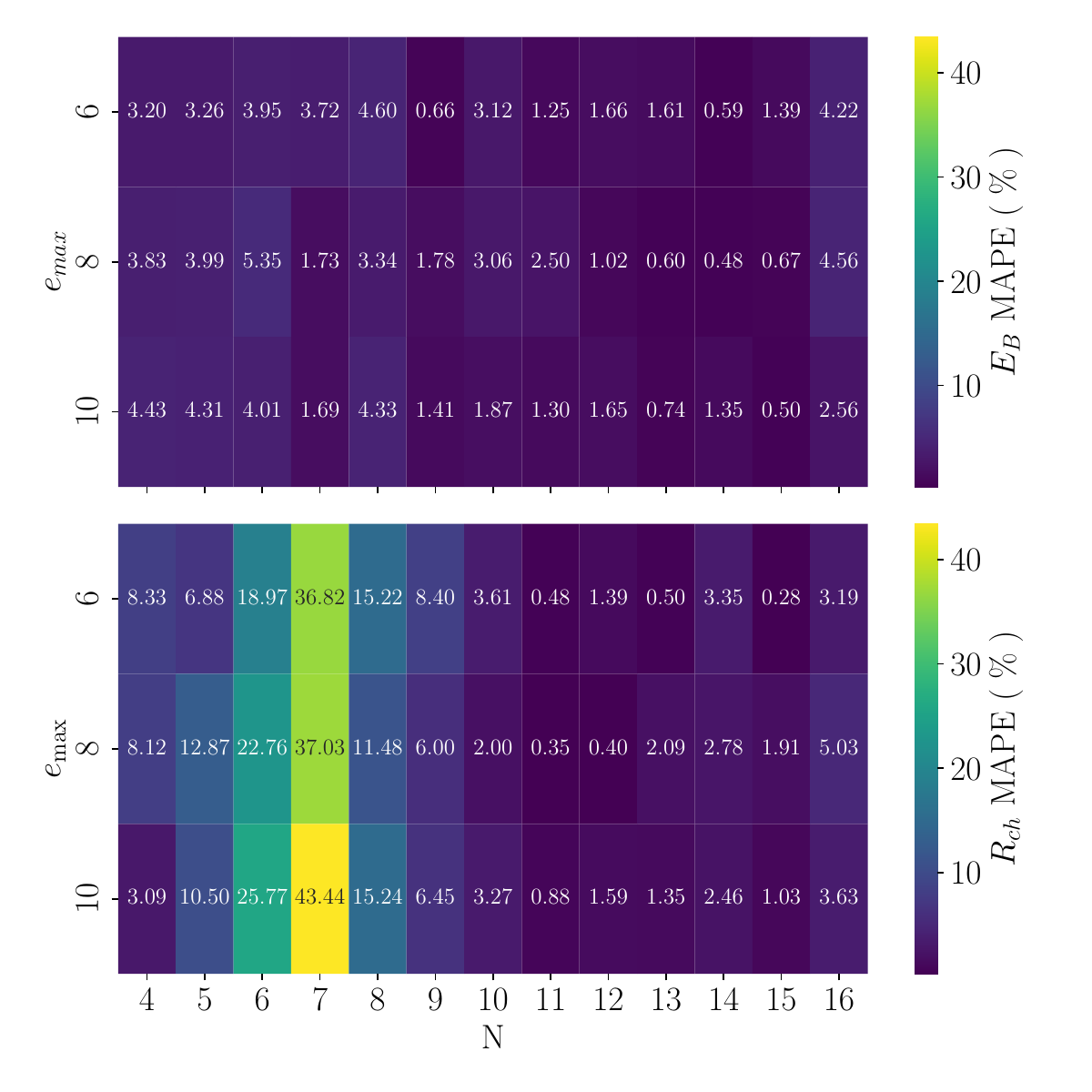}
    \caption{Low-Fidelity-Shot Leave-On-Out. Distribution of the $e_\text{max}$ extrapolation for unseen isotopes as data from $e_\text{max}>4$ was removed from the training Samples.}
    \label{fig:eemax_mape_heatmap}
\end{figure}

Following the same approach as with the Leave One Out - Zero Shot extrapolation presented above, we repeated this experiment but now including only data with $e_{\text{max}}=4$ for each one of the isotopes left out and the rest of the fidelities for the rest of the chain. In Fig.~\ref{fig:eemax_mape_heatmap} we show the evaluated MAPE as extrapolated.

\begin{table}[!htb]
    \caption{Inference ($e_{\text{max}}=4,6$) to high fidelity ($e_{\text{max}}=8$) with minimal extra data. Shown is the MAPE (\%) in binding energies and charge radii for oxygen isotopes after introducing a given fraction of the $e_{\text{max}}=8$ dataset.}
    \centering
    \begin{tabular}{c|c|c}
    \hline\hline
    \textbf{$e_{\text{max}}=8$ data $\%$} & \textbf{MAPE $B_E$ (\%)} & \textbf{MAPE $R_{ch}$ (\%)} \\
    \hline
    5\%  & 2.38 & 1.67 \\
    10\% & 1.65 & 1.24 \\
    20\% & 1.32 & 0.818 \\
    30\% & 1.52 & 0.557 \\
    40\% & 1.13 & 0.499 \\
    50\% & 1.16 & 0.362 \\
    60\% & 2.19 & 0.308 \\
    70\% & 0.650 & 0.326 \\
    80\% & 0.734 & 0.388 \\
    90\% & 0.571 & 0.267 \\
    \hline\hline
    \end{tabular}
    \label{tab:fine_tune}
\end{table}

Interestingly, adding the low-fidelity samples considerably helps reduce the MAPE, while difficulties in predicting the dynamics of LECs around the shell closure remain for the $R_{ch}$ regression.

\subsection{Data Efficiency}

Table~\ref{tab:fine_tune} summarizes the data usage and resulting RMSE, underscoring the computational savings and the robust uncertainty quantification still retained by the Bayesian framework. This model refinement significantly cuts computational cost: we can cheaply train the base model on large amounts of $e_{\text{max}}=4$ data and only sample a sparse set of high-$e_{\text{max}}$ which can be computed until the emulator residual converges.

\subsection{Ablation Study}

\begin{figure}[h]
    \centering
    \includegraphics[width=1\linewidth]{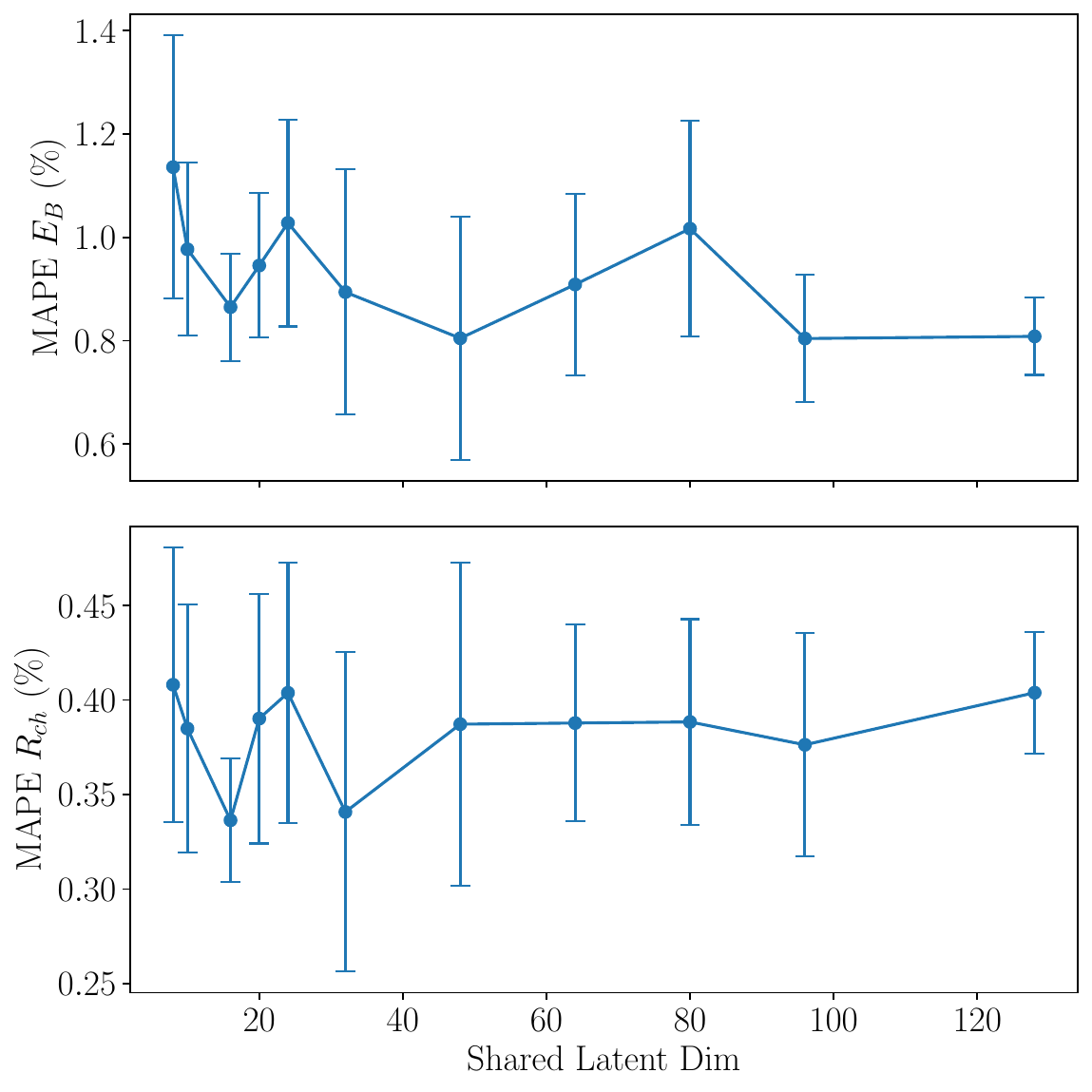}
    \caption{Ablation study of the size of the attention mechanism as we systematically increased the size of the Multihead Attention layer while keeping the other parameters fixed.
    \label{fig:ablation}}
\end{figure}

In order to model the behavior of the model as more attention-exclusive parameters are added, we performed a sweep among the shared dimension layer of the LEC before passing them to the attention mask, while keeping all of the other parameters of the model fixed. For each one of the data points we performed 10 trainings with different random seeds, and for each one of them evaluated the corresponding test samples. In Figure~\ref{fig:ablation}, the robustness of the interpolation performance is evident. The attention mechanism allows the model to capture higher-order correlations between the LECs, allowing for globally better emulation of both the binding energy and the charge radii. 

\subsection{General Sensitivity Analysis}

Global sensitivity analysis is a statistical tool used to decompose how the total uncertainty of a model is separated in term of the uncertainty of its different parameters~\cite{sobol2001global}. Following notation from \cite{Ekstrom:2019}, the total variance of the results, $Var[Y]$, is decomposed as
\begin{align}
    Var[Y] = \sum_{i=1}^{N_{LECs}} V_i +    \sum_j^{N_{LECs}}\sum_{i<j}^{N_{LECs}} V_{ij} + ... 
\end{align}
with the partial variances $V_i, V_{ij}, ...$ given by
\begin{align}
    &V_i= Var\left[E_{\vec{\alpha}\sim(\alpha_i)}[Y|\alpha_i]\right]\\
    &V_{ij}= Var\left[E_{\vec{\alpha}\sim(\alpha_i, \alpha_i)}[Y|\alpha_i, \alpha_j]\right] - V_i - V_j.
\end{align}
$\vec{\alpha}\sim(\alpha_i)$ is the set of LECs without $\alpha_i$ and $E_{\vec{\alpha}\sim(\alpha_i)}[Y|\alpha_i]$ is the conditional expectation of Y given $\alpha_i$, likewise for the higher order expression. We can then write the total sensitivity of the final results to each LEC $\alpha_i$ as 
\begin{align}
    S_T(\alpha_i) = S_i+ S_{ij} + S_{ijk} + ...
\end{align}
where the partial sensitivities are given by
\begin{align}
    S_i = \dfrac{V_i}{Var[Y]},  && S_{ij} = \dfrac{V_{ij}}{Var[Y]}
\end{align}
and likewise for higher orders.

\end{document}